\documentclass[12pt,aps,prb]{revtex4}%
\usepackage{amsmath}
\usepackage{amsfonts}
\usepackage{amssymb}
\usepackage{graphicx}%
\providecommand{\U}[1]{\protect\rule{.1in}{.1in}}
\begin{document}
\title{Relation between measurable and principal characteristics of radiation-induced
shape-change of graphite}
\author{M.V. Arjakov}
\author{A.V. Subbotin}
\email{subbotin@atomtechnoprom.ru}
\affiliation{Scientific and Production Complex Atomtekhnoprom, Moscow, 119180, Russia}
\author{S.V. Panyukov}
\email{panyukov@lpi.ru}
\author{O.V. Ivanov}
\email{ivanov@lpi.ru}
\affiliation{P.N. Lebedev Physics Institute, Russian Academy of Sciences, Moscow, 117924, Russia}
\author{A.S. Pokrovskii}
\author{D.V. Kharkov}
\affiliation{State Scientific Center of the Russian Federation - Research Institute for
Nuclear Reactors, Dimitrovgrad, 433510, Russia}

\begin{abstract}
On the basis of studies of radiation-induced shape-change of reactor graphite
GR-280, through the series of measurements of samples with different
orientation of cutting with respect to the direction of extrusion, a
conclusion is made about the existence of polycrystal substructural elements
-- domains. Domains, like graphite as a whole, possess the property of
transverse isotropy, but have different amplitudes of shape-change and random
orientations of the axes of axial symmetry. The model of graphite, constructed
on the basis of the concept of domains allowed to explain from a unified point
of view most of existing experimental data. It is shown that the presence of
the disoriented domain structure leads to the development of radiation-induced
stresses and to the dependence of the shape-change on the size of graphite
samples. We derive the relation between the shape-change of finite size
samples and the actual shape-change of macro-graphite.

\end{abstract}
\maketitle

\newpage

\section{Introduction}

Reactor graphites of various grades, used in nuclear engineering as moderators
of neutrons and constructional materials, are subjected to high dose damaging
radiation. At that, developing in graphites radiation-induced physical,
mechanical and dimensional effects significantly affect their operational characteristics.

Detailed analysis of such irradiation effects as a shape-change, evolution
with irradiation dose of elastic moduli, thermal expansion coefficient,
temperature- and electrical conductivity, radiation creep, etc., suggests an
essential role of graphite morphology in all irradiation
effects.\cite{n+1,n+2,n+3,n+4,n+4a} The morphology of graphite is formed by
two main components -- the filler and the binder, as well as due to the
presence of an ensemble of microcracks and technological
pores\cite{n+1,n+4,n+4a}, and to a large extent is determined by the
technology of graphite fabrication. The binder has a fine crystalline
structure, whereas the structure of the filler is hierarchical, based on
microcrystallites with more or less perfect hexagonal crystal lattice of two
types -- $ABAB...$ and $ABCABC$, the fraction of which may be
different.\cite{n} Between the crystallite with hexagonal structure and the
macro-graphite as a whole (meaning the graphite as a material of finite sizes)
there is a succession of scaled structural levels of organization of
crystallites, of a binder and ensembles of microcracks, which form the
morphology of graphite.\cite{n+4a,n+5}

Elastic constants of hexagonal crystallographic lattice are described by five
independent constants $c_{11},c_{12},c_{13},c_{33}$ and $c_{44}$.\cite{n+6} It
is established that reactor graphites, obtained by extruding or pressing, have
properties of a transversely isotropic medium\cite{n+1,n+7}, and to describe
their elasticity they also require five independent variables, for example,
$E_{\perp},E_{\parallel},G_{\parallel},\nu_{\perp},\nu_{\parallel}$ -- Young's
and shear moduli and Poisson's ratio in the isotropy plane and normal to it,
respectively. Radiation-induced shape-change of graphite has similar symmetry
properties, allowing us to introduce the diagonal tensor of radiation-induced
shape-change (see below).

For extruded graphites the axis of axial symmetry is given by the direction of
extrusion ($\parallel$ direction), whereas directions normal to it ($\perp$
directions) define the plane of isotropy.\cite{n+1} The well known reactor
graphite GR-280\cite{n+4a} belongs to this group of graphites. In the present
paper we discuss experimental results obtained from studies of
radiation-induced shape-change of graphite GR-280.

\section{Description of experiment}

Graphite samples in an amount of several hundreds obtained by longitudinal and
transverse cut from the bulk material ($\parallel$ and $\perp$ orientations),
of diameters $6$ mm, $8$ mm, $12$ mm, and of length $50$ mm were exposed to
irradiation at temperatures $460\pm25^{\circ}$C, $550\pm25^{\circ}$C,
$640\pm25^{\circ}$C in the research fast-breeder reactor BOR-60 up to dose
levels $\sim3\times10^{22}n/cm^{2}$ at energies $E_{n}>0,18$ Mev. Measurements
were carried out with the dose intervals $\sim0.6\times10^{21}n/cm^{2}$. The
samples were measured both in longitudinal and transverse directions, using
the $4.5$ mm step. When measuring the cross-section the diameters at $32$
points on the circle were recorded.

All data are presented as functions of doses of neutrons with the spectrum of
BOR-60. Dose conversion to the neutron spectra of thermal reactors are
presented in Ref.\cite{n+9}. The conversion of neutron doses BOR-60 into
equivalent doses and fluxes of neutrons of thermal reactors requires a special
consideration and will be given in a separate paper.

As the result of mass measurements of samples, in the study of patterns of
their shape-change the two observations were obtained, that required an
additional explanation:

\begin{itemize}
\item Circular cross-section of samples acquires with the irradiation dose an
increasingly pronounced elliptical shape, and orientations of ellipses vary
randomly for measurements along the sample on the scale of $\sim6$ mm.

\item Results of calculation of the relative volume changes using conventional
expression\cite{n+1,n+2,n+3}
\begin{equation}
\left(  \Delta V/V\right)  ^{C}=\left(  \Delta L/L\right)  ^{\parallel
}+2\left(  \Delta L/L\right)  ^{\perp}, \label{dV/V}%
\end{equation}
(where $\left(  \Delta L/L\right)  ^{\parallel}$ and $\left(  \Delta
L/L\right)  ^{\perp}$ are relative length changes of the samples with parallel
and perpendicular orientations, respectively) increasingly diverge with
irradiation dose from results of direct measurements of the relative change of volume.
\end{itemize}

Below we present the interpretation of the above facts in the framework of the
model evolved by authors that takes into account hierarchical morphology of
graphite.\cite{1} Defining the characteristic element of a certain scale in
the scaling hierarchy as radiation-induced affine shape-changing region, the
concept of the \textquotedblleft domain\textquotedblright\ is introduced, as a
region of intermediate size between the grain of graphite and the
macro-graphite (of size $\sim5$ mm). The domain has the same symmetry (but
different values) of radiation-induced shape change as macro-graphite,
nevertheless its symmetry axis does not coincide with the axis of extrusion.
We use this concept to develop a formalism that allows to bind together all
the above facts.

\section{Processing experimental data}

In this section we describe the geometry of graphite samples before and after
irradiation and introduce the main physical concepts to describe
radiation-induced change of their shape. Using these concepts we find the
relation between local characteristics of domain structure and experimentally
observed radiation-induced shape-changes of graphite samples.

\begin{figure}
[tbh]
\begin{center}
\includegraphics[
height=2.2795in,
width=4.8866in
]%
{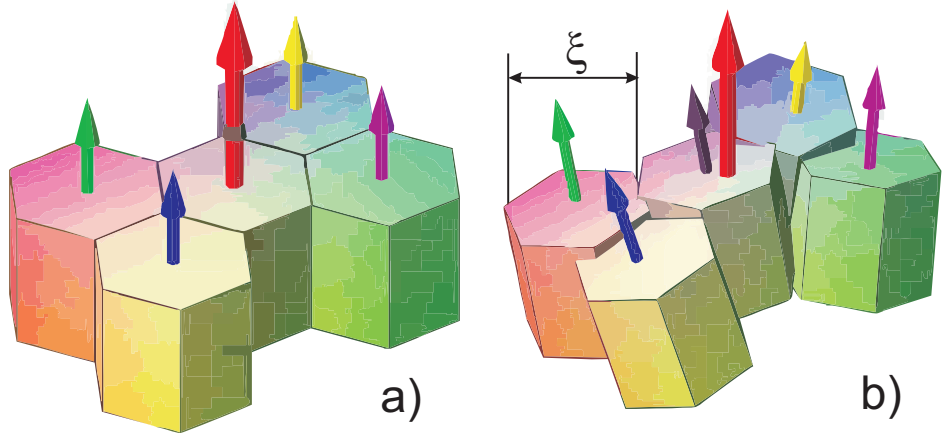}%
\caption{Graphite is a polycrystal consisting of domains (shown schematically
by hexahedrons) of typical size $\xi$. Each domain can be characterized by the
local direction of anisotropy (shown by small arrows). a) In the case of
homogeneous structure all domains have the same orientation. b) In the case of
heterogeneous structure these directions are random vectors with preferential
orientation (shown by large arrow), determined by the specimen orientation.
Radiation-induced shape-change of randomly oriented domains leads to the rise
of cross-domain stresses with irradiation doze.}%
\label{Domains}%
\end{center}
\end{figure}

\emph{\textbf{Domains.}} Graphite is a polycrystal, and as each polycrystal it
consists of small microcrystallites. The most important difference of graphite
from usual polycrystals is the presence of mesoscale three-dimensional objects
with the size $\xi\gtrsim0.5$ cm for brevity called domains, see
Fig.~\ref{Domains}. Each domain consists of many microcrystallites, possesses
the properties of transversely isotropic medium and can be characterized by
individual orientation of the axis of symmetry. At preparation of graphite the
local orientations of anisotropy axes of different domains deviate from their
average over the sample orientation due to technological reasons. The change
of shape of domains under irradiation is restricted by their local
environment, since this change is constrained by neighboring microcrystallites
of different orientations, leading to the development of internal stresses in
the bulk material, see Fig.~\ref{Domains} b.
These stresses lead to the additional strong dimension effect in graphite. The
knowledge of the dimension effect is extremely important to understand the
behavior of bulk graphite materials, whereas most of experimental data are
obtained for finite size graphite samples.

Because of transversal isotropy of a domain its deformation can be described
by two principal strains: along the principal axis of axial anisotropy,
$\left(  \Delta l/l\right)  _{\parallel}$, and perpendicular to it, $\left(
\Delta l/l\right)  _{\perp}$. These strains determine the tensor of
radiation-induced deformation $\hat{F}_{0}$ which is diagonal in local
coordinate system, axes of which are directed along the main axes of local
deformations:%
\begin{equation}
\hat{F}_{0}=\left(
\begin{array}
[c]{ccc}%
1+\left(  \Delta l/l\right)  _{\perp} & 0 & 0\\
0 & 1+\left(  \Delta l/l\right)  _{\perp} & 0\\
0 & 0 & 1+\left(  \Delta l/l\right)  _{\parallel}%
\end{array}
\right)  . \label{F0}%
\end{equation}

\begin{figure}
[tbh]
\begin{center}
\includegraphics[
height=2.4835in,
width=5.0508in
]%
{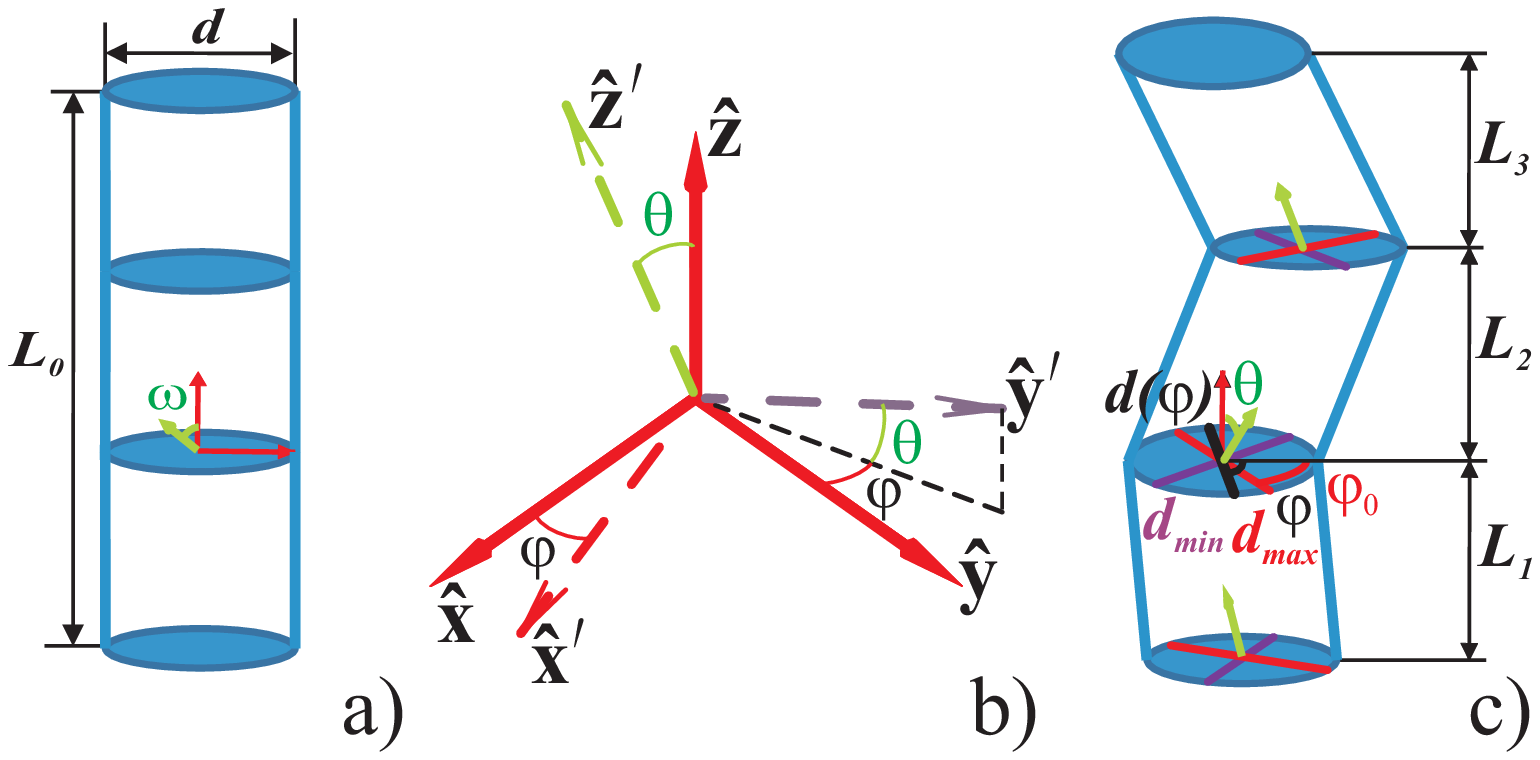}%
\caption{Before irradiation the sample has cylindrical shape with circular
cross-sections, Figure a). Different coordinate systems b): $\left(
\mathbf{\hat{x}},\mathbf{\hat{y}},\mathbf{\hat{z}}\right)  $ is attached to
the sample with axis $\mathbf{\hat{z}}$ along the preferential direction,
$\left(  \mathbf{\hat{x}}^{\prime},\mathbf{\hat{y}}^{\prime},\mathbf{\hat{z}%
}^{\prime}\right)  $ is related to a domain with axis $\mathbf{\hat{z}%
}^{\prime}$ along the direction of axial symmetry of the domain. After
irradiation the sample takes a \textquotedblleft crankshaft\textquotedblright%
\ shape with elliptical cross-sections, characterized by maximal $d_{\max}$
and minimal $\left(  d_{\min}\right)  _{k}$ diameters measured along main axes
rotated by random angles $\left(  \varphi_{0}\right)  _{k}$, Figure c).}%
\label{Fig-1}%
\end{center}
\end{figure}

\emph{\textbf{a) Homogeneous structure.}} Before irradiation the specimen has
the shape of a cylinder of length $L$ and diameter $d$. After irradiation the
shape of the cylinder will be changed because of radiation induced
shape-change of randomly oriented domains. We first consider the case when
orientations of axial symmetry axes of all domains of the sample coincide with
the extrusion direction, so every sample can be characterized by the
orientation angle $\omega$ between the direction of axial symmetry and the
axis of the specimen, see Figs.~\ref{Domains} a and~\ref{Fig-1} a. Although
the approximation of homogeneous texture can be justified only for small
disorder in domain orientation, it is commonly used in the literature because
of the simplicity of measurements of sample sizes.%

Relative deformations of the specimens along $i=x,y,z$ -directions in the
coordinate system related to the sample are described by the deformation
tensor $\hat{F}_{ideal}$, $\left(  \Delta L_{i}/L_{i}\right)  _{ideal}=\left(
\hat{F}_{ideal}\right)  _{ii}-1$. In the case of homogeneous texture the
deformation tensor has the form
\begin{equation}
\hat{F}_{ideal}=\hat{R}_{1}\left(  \omega\right)  \hat{F}_{0}\hat{R}_{1}%
^{-1}\left(  \omega\right)
\end{equation}
where $\hat{R}_{1}\left(  \omega\right)  $ is the tensor of rotation by the
orientation angle $\omega$:%
\begin{equation}
\hat{R}_{1}\left(  \omega\right)  =\left(
\begin{array}
[c]{ccc}%
1 & 0 & 0\\
0 & \cos\omega & -\sin\omega\\
0 & \sin\omega & \cos\omega
\end{array}
\right)  . \label{rot1}%
\end{equation}

\emph{\textbf{b) The structure with random domain orientations.}} The
orientation of local axial axis in a cylindrical specimen can be characterized
by two angles: the azimuth angle $\varphi_{0}$ in the cross-sectional plane
$\left(  x,y\right)  $ and the polar angle $\theta$ between the local axial
direction and the cylinder axis, see Fig.~\ref{Fig-1} b. Both these angles
randomly vary in space, and their average values, $\bar{\varphi}_{0}$ and
$\omega=\overline{\theta}$ depend on orientation of the sample -- how it was
cut from a bulk graphite array. In the following we will consider both
parallel (with average axial direction along the cylinder axis, $\omega
=0^{\circ}$, superscript $\parallel$) and perpendicular (with average axial
direction perpendicular to the cylinder axis, $\omega=90^{\circ}$, superscript
$\perp$) orientations.

Macroscopic deformation of the sample is described by deformation gradient
tensor $\hat{F}$ defined in coordinate system, related to the specimen. For
the axial axis rotated by the angles $\left(  \varphi_{0},\theta\right)  $
with respect to the axis of the cylinder the gradient tensor takes the form
\begin{equation}
\hat{F}=\hat{R}\left(  \varphi_{0},\theta\right)  \hat{F}_{0}\hat{R}%
^{-1}\left(  \varphi_{0},\theta\right)  \label{T}%
\end{equation}
Here $\hat{R}\left(  \varphi_{0},\theta\right)  \mathbf{=}\hat{R}_{2}\left(
\varphi_{0}\right)  \hat{R}_{1}\left(  \theta\right)  $ and the tensor of
rotation in the plane of cross-section of the sample $\hat{R}_{2}\left(
\varphi_{0}\right)  $ is defined by expression%
\begin{equation}
\hat{R}_{2}\left(  \varphi_{0}\right)  =\left(
\begin{array}
[c]{ccc}%
\cos\varphi_{0} & -\sin\varphi_{0} & 0\\
\sin\varphi_{0} & \cos\varphi_{0} & 0\\
0 & 0 & 1
\end{array}
\right)  . \label{rot2}%
\end{equation}
The orientation of the deformation tensor $\hat{F}$ of a domain randomly
varies in space on the length $\xi$ about the domain size. As the result of
such variations of $\hat{F}$ the specimen is deformed after irradiation into a
\textquotedblleft crankshaft\textquotedblright\ shape, see Fig.~\ref{Fig-1} c.
The bending of the cylindrical specimen at locations of domain junctions is
described by effects of higher order in angle deviations and will not be
studied below, although this effect is steadily fixed experimentally. Consider
the shape of the crankshaft cylinder in more details:

\emph{\textbf{Transverse dimensions.}} Before irradiation the cross-section of
the specimen has circular shape with angular dependence of the diameter
$\mathbf{d}_{0}\left(  \varphi\right)  =\left(  d\cos\varphi\right)
\mathbf{\hat{x}}+\left(  d\sin\varphi\right)  \mathbf{\hat{y}}$. After
irradiation the shape $\mathbf{d}\left(  \varphi\right)  $ of the
cross-section is determined by the projection of the vector $\hat{F}%
\mathbf{d}_{0}\left(  \varphi\right)  $ on the cross-sectional plane $\left(
\mathbf{\hat{x}},\mathbf{\hat{y}}\right)  $. Calculating the square of the
cross-sectional diameter in the deformed sample we find its dependence on
azimuth angle $\varphi$:%
\begin{equation}
d^{2}\left(  \varphi\right)  =d_{\max}^{2}\cos^{2}\left(  \varphi-\varphi
_{0}\right)  +d_{\min}^{2}\sin^{2}\left(  \varphi-\varphi_{0}\right)
\label{dphi}%
\end{equation}
Thus, initially circular cross-section of diameter $d$ after irradiation takes
elliptical shape, see Fig.~\ref{Fig-1} c. Maximal $d_{\max}$ and minimal
$d_{\min}$ diameters of the ellipse are given by equations:%
\begin{align}
d_{\max}/d  &  =1+\left(  \Delta l/l\right)  _{\perp},\label{dmax}\\
d_{\min}/d  &  =1+\left(  \Delta l/l\right)  _{\perp}\cos^{2}\theta+\left(
\Delta l/l\right)  _{\parallel}\sin^{2}\theta\label{dmin}%
\end{align}
According to Eq.~(\ref{dmax}) the maximal diameter $d_{\max}$ is identical for
all segments of the specimen, while minimum diameter $d_{\min}$ randomly
varies between different segments.

\emph{\textbf{Relative elongation.}} We denote by $L_{0k}$ the initial length
of $k$-th domain. After irradiation the length $L_{k}$ of the $k$-th segment
of the crankshaft cylinder can be found as the projections of the vector
$\hat{F}\left(  L_{0k}\mathbf{\hat{z}}\right)  $ on the direction
$\mathbf{\hat{z}}$ of unit vector along the axis of the cylinder:%
\begin{equation}
L_{k}=L_{0k}\left[  1+\left(  \Delta l/l\right)  _{\perp}\sin^{2}%
\theta+\left(  \Delta l/l\right)  _{\parallel}\cos^{2}\theta\right]  \label{L}%
\end{equation}
The total length of the specimen is found as the sum of lengths of all its
segments%
\begin{equation}
L=\sum\nolimits_{k=1}^{N}L_{k}\label{Lk}%
\end{equation}
From Eqs.~(\ref{Lk}) we get the relative change of the length%
\begin{equation}
\left(  \Delta L/L\right)  =\left(  \Delta l/l\right)  _{\perp}\overline
{\sin^{2}\theta}+\left(  \Delta l/l\right)  _{\parallel}\overline{\cos
^{2}\theta}=\left(  \Delta l/l\right)  _{\parallel}+\bar{\varepsilon
}\label{dLe}%
\end{equation}
Here $\left(  \Delta L/L\right)  $ is measured as relative elongation of the
whole length of the specimen, while $\left(  \Delta l/l\right)  _{\perp}$ and
$\left(  \Delta l/l\right)  _{\parallel}$ are corresponding components of the
principal strain. We use the notation $\overline{\cdots}$ for averaging over
the sample:%
\begin{equation}
\bar{\varepsilon}=\frac{1}{N}\sum\limits_{k=1}^{N}\varepsilon_{k}=\left[
\left(  \frac{\Delta l}{l}\right)  _{\perp}-\left(  \frac{\Delta l}{l}\right)
_{\parallel}\right]  \overline{\sin^{2}\theta},\label{eav}%
\end{equation}
and $\varepsilon_{k}$ is the flattening factor of $k$-th cross-section:%
\begin{equation}
\varepsilon_{k}=\frac{d_{\max}-\left(  d_{\min}\right)  _{k}}{d}=\frac{\Delta
d_{\max}-\left(  \Delta d_{\min}\right)  _{k}}{d}\label{eps}%
\end{equation}

\emph{\textbf{Volume change.}} The volume of the specimen can be found as the
sum of volumes of all its segments, see Fig.~\ref{Fig-1} c:%
\begin{equation}
V=\sum\nolimits_{k=1}^{N}\frac{\pi}{4}d_{\max}\left(  d_{\min}\right)
_{k}L_{k}\simeq\frac{\pi}{4}d_{\max}\overline{d_{\min}}L \label{Vd}%
\end{equation}
Here $\frac{\pi}{4}d_{\max}\left(  d_{\min}\right)  _{k}$ is the area of
$k$-th cross-section. Using this equation we get the relative change of the
volume of such \textquotedblleft free\textquotedblright\ sample%
\begin{align}
\left(  \Delta V/V\right)  ^{F}  &  =\left(  \Delta d_{\max}/d\right)
+\left(  \overline{\Delta d_{\min}}/d\right)  +\Delta L/L\nonumber\\
&  =2\left(  \Delta d_{\max}/d\right)  +\Delta L/L-\bar{\varepsilon}
\label{dVF}%
\end{align}
According to Eqs.~(\ref{dmax}) and~(\ref{dLe}) the volume change can be
expressed through the principal strains of the domain:%
\begin{equation}
\left(  \Delta V/V\right)  ^{F}=Tr\left(  \hat{F}_{0}-\hat{1}\right)
=2\left(  \Delta l/l\right)  _{\perp}+\left(  \Delta l/l\right)  _{\parallel}
\label{dVV}%
\end{equation}
where $Tr\left(  \cdots\right)  $ is the sum of diagonal elements.
Eq.~(\ref{dVV}) demonstrates the self-consistency of our definition of
principal strains, since it reproduces the experimentally measurable volume change.

\emph{\textbf{Processing experimental data.}} In the following we will
calculate the two principal strains $\left(  \Delta l/l\right)  _{\parallel}$
and $\left(  \Delta l/l\right)  _{\perp}$ through the change of experimentally
measurable maximal diameter and length of the specimen:
\begin{equation}
\left(  \Delta l/l\right)  _{\parallel}=\Delta L/L-\bar{\varepsilon}%
,\qquad\left(  \Delta l/l\right)  _{\perp}=\Delta d_{\max}/d \label{dll}%
\end{equation}
where we used Eqs.~(\ref{dLe}) and~(\ref{dmax}). In our experiments the set of
diameters $\left(  d_{i}\right)  _{k}$ of the specimen is measured along $N$
equally separated cross-sections $k$ for $M$ equidistant angles $\varphi
_{j}=2\pi j/M$. Each of these sections can be characterized by minimal
diameter of the ellipsis $\left(  d_{\min}\right)  _{k}$ and the angle
$\left(  \varphi_{0}\right)  _{k}$ of its rotation in the cross-sectional
plane, while the maximum diameter $d_{\max}$ of the ellipse is the same for
the whole sample. The values of these parameters are found from experimental
set of diameters $\left\{  \left(  d_{i}\right)  _{k}\right\}  $ minimizing
the mean squared deviations
\[
\sigma^{2}=\sum\nolimits_{k=1}^{N}\sum\nolimits_{i=1}^{M}\left[  \left(
d_{i}^{2}\right)  _{k}-d_{k}^{2}\left(  \varphi_{i}\right)  \right]  ^{2}%
\]
with respect to $d_{\max}$ and the set of $N$ individual diameters $\left(
d_{\min}\right)  _{k}$ and angles $\left(  \varphi_{0}\right)  _{k}$ for each
of $N$ sections. Here $d_{k}^{2}\left(  \varphi\right)  $ is given by
Eq.~(\ref{dphi}) with corresponding parameters $\left(  d_{\min}\right)  _{k}$
and $\left(  \varphi_{0}\right)  _{k}$. The solution of these minimum
conditions is quite cumbersome and rendered in Appendix. Here we show the
result%
\begin{align}
d_{\max}^{2}  &  =\frac{1}{N}\sum\nolimits_{k=1}^{N}\left(  A_{k}+\sqrt
{B_{k}^{2}+C_{k}^{2}}\right)  ,\label{Dmax}\\
\left(  d_{\min}^{2}\right)  _{k}  &  =\frac{4}{3}A_{k}-\frac{2}{3}\sqrt
{B_{k}^{2}+C_{k}^{2}}-\frac{1}{3}d_{\max}^{2},\label{Dmin}\\
\left(  \varphi_{0}\right)  _{k}  &  =\frac{1}{2}\arctan\frac{C_{k}}{B_{k}},
\label{phik}%
\end{align}
where $A_{k},B_{k}$ and $C_{k}$ are Fourier transforms of squared diameters of
the specimens:
\begin{align}
A_{k}  &  =\frac{1}{M}\sum\nolimits_{i=1}^{M}\left(  d_{i}^{2}\right)
_{k},\label{Ak}\\
B_{k}  &  =\frac{2}{M}\sum\nolimits_{i=1}^{M}\left(  d_{i}^{2}\right)
_{k}\cos\left(  2\varphi_{i}\right)  ,\label{Bk}\\
C_{k}  &  =\frac{2}{M}\sum\nolimits_{i=1}^{M}\left(  d_{i}^{2}\right)
_{k}\sin\left(  2\varphi_{i}\right)  . \label{Ck}%
\end{align}

\section{Interpretation of data}

In this section we present observable dependences of geometrical sizes of
samples on fluence $\Phi$. In the following we show only curves averaged over
the row of specimens with the same orientation ($\parallel$ and $\perp$),
diameter ($6$ and $8$ mm) and average irradiation temperature $T$. Samples of
assembly $A$ were exposed to radiation in reactor Bor-60 at the temperature
$T=460\pm25^{\circ}$C, samples of assemblies $B$ and $C$ have irradiation
temperatures $550\pm25^{\circ}$C and $640\pm25^{\circ}$C, respectively.

\begin{figure}
[tbh]
\begin{center}
\includegraphics[
height=3.0583in,
width=5.4777in
]%
{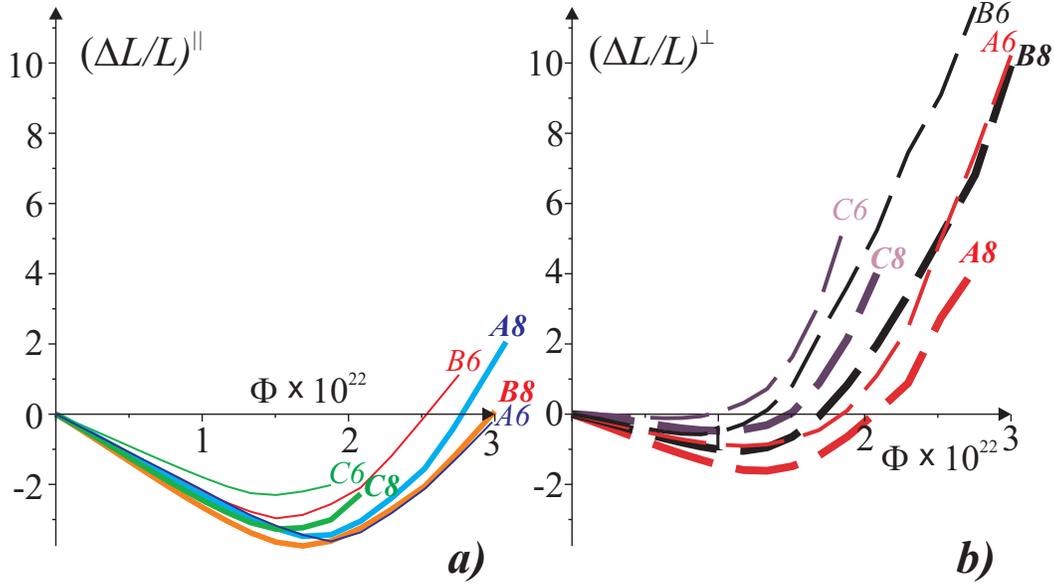}%
\caption{The dependence of relative length changes (in \%) for samples $(ad)$
from the assembly $a=A,B,C$ with diameter $d$ ($6$ mm - thin lines, $8$ mm -
thick lines) with parallel ($\parallel$) a) and perpendicular ($\perp$) b)
orientation $\omega$ on neutron fluence $\Phi$ ($E_{n}>0.18$ Mev).}%
\label{Length}%
\end{center}
\end{figure}

\emph{\textbf{a) Length changes.}} Most information about radiation-induced
shape-change of graphite is usually obtained using data describing the change
of length $L$ of samples with neutron fluence $\Phi$. For example, the volume
change $\Delta V$ is commonly estimated by the formula~(\ref{dV/V}), using the
data on length variations for specimens of parallel, $\left(  \Delta L\right)
^{\parallel}$, and perpendicular, $\left(  \Delta L\right)  ^{\perp}$,
orientations. The dependence of relative elongations $\left(  \Delta
L/L\right)  ^{\parallel}$ and $\left(  \Delta L/L\right)  ^{\perp}$ of studied
samples on fluence $\Phi$ is shown in Fig.~\ref{Length}.%
As one can see, the change of the sample size strongly depends on sample shape
because of different constraints imposed on domains changing their shape in
samples of different diameter $d$ and orientation $\omega$. The radiation
induced shape-change of domains is only weakly restricted in specimens with
perpendicular orientation and of smallest diameter $d=6$ mm, while the domains
experience strongest restrictions in specimens with parallel orientation and
of diameter $d=8$ mm$.$

\begin{figure}
[tbh]
\begin{center}
\includegraphics[
height=2.0359in,
width=5.5372in
]%
{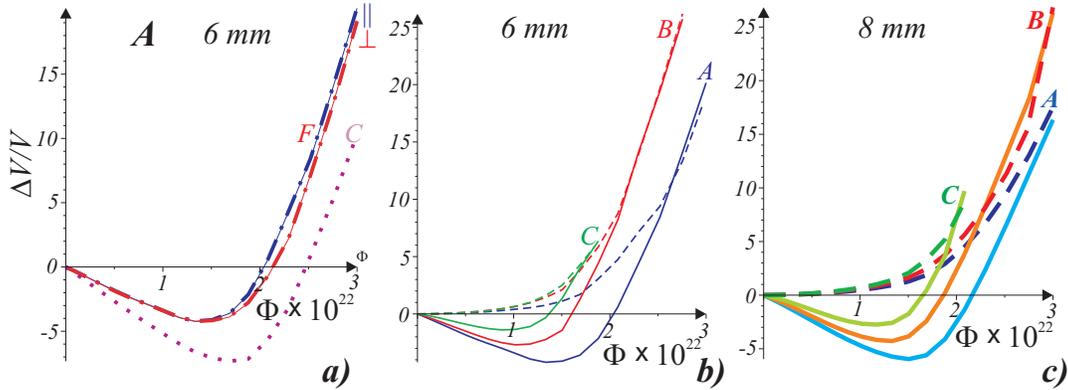}%
\caption{Relative volume change $\Delta V/V$ (in \%) as function of fluence
$\Phi$ for assembly $A$ with temperature $T=460\pm25^{\circ}$C. a) solid lines
($F$) show $\left(  \Delta l/l\right)  _{\parallel}+2\left(  \Delta
l/l\right)  _{\perp}$, dash-dotted lines show $\Delta L/L+2\left(  \Delta
d/d\right)  $ for $6$ mm samples of both parallel ($\parallel$) and
perpendicular ($\perp$) orientations. Since similar curves are obtained for
all other samples, in figures b) and c) we show only single dependence of
$\Delta V/V=\Delta L/L+2\left(  \Delta d/d\right)  $ on $\Phi$ for assemblies
$A,B$ and $C$ (solid lines, with average temperatures $460,550$ and
$640^{\circ}$C, respectively), and also asymptotic dependence of these curves
$C[\left(  \Delta l/l\right)  _{\perp}-\left(  \Delta l/l\right)  _{\parallel
}]^{2}$ (dashed lines, $C\simeq7$ for all curves) at large fluence $\Phi$.}%
\label{Volume}%
\end{center}
\end{figure}

\emph{\textbf{b) Volume changes}}. In Fig.~\ref{Volume} we show the dependence
of the sample volume on fluence $\Phi$.%

The evolution of the sample volume with the fluence $\Phi$ can be calculated
in two different ways: using variations of macroscopic sample sizes
(Eq.~(\ref{dVF})) and using the change of local sizes of domains
(Eq.~(\ref{dVV})). As one can see from figure~\ref{Volume} a), both ways give
essentially the same result:
\begin{equation}
\left(  \Delta V/V\right)  ^{F}=\Delta L/L+2\left(  \Delta d/d\right)
=\left(  \Delta l/l\right)  _{\parallel}+2\left(  \Delta l/l\right)  _{\perp}
\label{Vt}%
\end{equation}
We observe that the volume of the specimen only slightly varies with its
orientation $\omega$. Therefore, values of principal strains $\left(  \Delta
l/l\right)  _{\parallel}$ and $\left(  \Delta l/l\right)  _{\perp}$, through
which the volume change is expressed, can be considered as universal
characteristic of graphite, weakly depending on the specimen orientation
$\omega$. In contrast to the volume change (see Fig.~\ref{Volume}) the length
$L$ and the diameter $d$ of the sample (see Fig.~\ref{Length}) are not
universal and strongly depend on the sample shape. These values depend not
only on principal strains, but also on the flattening factor $\varepsilon$
which is not universal because of different constraints imposed on domains
changing their shape under irradiation in small specimens of different
orientations. We observe that variations of the volume are depressed for
samples of larger diameter, $d=8$ mm, because of stronger constraints imposed
on shape-changing domains in larger samples.

The dependence of volume change on fluence $\Phi$ is strongly non-monotonic:
at small fluences $\Phi<\Phi_{\min}$ the graphite shrinks with dose $\Phi$,
and it dilates at $\Phi>\Phi_{\min}$. Dotted lines in Figs.~\ref{Volume} a)
and b) show the dependence%
\begin{equation}
C\left[  \left(  \Delta l/l\right)  _{\perp}-\left(  \Delta l/l\right)
_{\parallel}\right]  ^{2} \label{asimp}%
\end{equation}
with coefficient $C=7$. As one can see, this dependence describes relatively
well asymptotic behavior at large $\Phi>\Phi_{\min}$ of volume changes in
graphite for all temperatures, diameters and orientations. The knowledge of
the asymptotic behavior is important both to understand undergoing mechanisms
of graphite deformation under irradiation and to extrapolate experimental data
in a region of higher radiation dozes not reachable for current experiments.

Using Eq.~(\ref{dLe}) we find the following expression for combined
equation~(\ref{dV/V}) traditionally used to describe the volume change $\Delta
V^{C}$ during irradiation:%
\begin{equation}
\left(  \Delta V/V\right)  ^{C}=\left(  \Delta V/V\right)  ^{F}-\left(
2\bar{\varepsilon}^{\perp}-\bar{\varepsilon}^{\parallel}\right)  <\left(
\Delta V/V\right)  ^{F} \label{Vprime}%
\end{equation}
We observe, that Eq.~(\ref{Vprime}) gives different results ($\Delta
V^{C}<\Delta V^{F}$) from real relative change of sample volume given by
Eq.~(\ref{Vt}) (see dotted line in figure~\ref{Volume}) due to ellipticity,
$\bar{\varepsilon}>0$, of specimen cross-sections. We conclude that the
traditional expression~(\ref{Vprime}) can not be used to calculate the actual
change of volume of graphite samples. The replacement of actual volume change
$\Delta V^{F}$ by $\Delta V^{C}$ can be safety applied only in the case of
homogeneous texture of graphite, see Fig.~\ref{Domains} a. Disorder in domain
orientations (Fig.~\ref{Domains} b) leads to significant difference between
commonly used (Eq.~(\ref{Vprime})) and actual (Eq.~(\ref{Vt})) volumes of samples.

\emph{\textbf{c) Parameters of domains.}} The orientation of domain anisotropy
axis is determined by two angles: the azimuth angle $\varphi_{0}$ and the
polar angle $\theta$, see Fig.~\ref{Fig-1} b). Below we estimate parameters of
domains from the knowledge of statistics of these angles for samples under consideration.

Azimuth angle $\varphi_{0}$ determines orientation of the axial axis in the
cross-sectional plane. The angles $\left(  \varphi_{0}\right)  _{k}$ are
random variables of the cross-section number $k$, that are correlated at the
distance about the domain size $\xi$. In general, the amplitude of angle
fluctuations can be different for different samples. In order to exclude the
effect of such amplitude variations consider the ratio%

\begin{equation}
r=\frac{2r_{g}}{Nr_{n}} \label{r}%
\end{equation}
of two sums for each of samples
\begin{equation}
r_{g}=\sum_{k=1}^{N}\sum_{j=k+1}^{N}\left[  \left(  \varphi_{0}\right)
_{k}-\left(  \varphi_{0}\right)  _{j}\right]  ^{2},\qquad r_{n}=\sum
_{k=1}^{N-1}\left[  \left(  \varphi_{0}\right)  _{k}-\left(  \varphi
_{0}\right)  _{k+1}\right]  ^{2} \label{Cor}%
\end{equation}
The first sum is going over all pairs of cross-sections (an analog of a
gyration radius), while the second one is only over nearest pairs (an analog
of end-to-end distance of a polymer chain). Since each of these terms is
quadratic form of angles $\left(  \varphi_{0}\right)  _{k}$, their
ratio~(\ref{r}) does not depend on amplitude of fluctuations and encrypts only
information about domain size. If there are no correlations between angles
$\varphi_{k}$ this ratio tends to $1$ in the limit of large $N\rightarrow
\infty$, when we can change $[\left(  \varphi_{0}\right)  _{k}-\left(
\varphi_{0}\right)  _{j}]^{2}$ in Eqs.~(\ref{Cor}) by its average,
$\overline{[\left(  \varphi_{0}\right)  _{k}-\left(  \varphi_{0}\right)
_{j}]^{2}}=2\overline{\varphi_{0}^{2}}$. In order to determine the dependence
of the parameter $r$ on domain size $\xi$, consider the case when the group of
$l$ consecutive angles $\varphi_{nl+j}$, $j=1,\ldots,l$ has the same value
$\hat{\varphi}_{n}$ for integer $n,l$, whereas the values $\hat{\varphi}_{n}$
for each group $n$ are independent random variables. At large $N\rightarrow
\infty$ changing $[\left(  \varphi_{0}\right)  _{k}-\left(  \varphi
_{0}\right)  _{j}]^{2}$ in Eq.~(\ref{Cor}) by its average ($0$ for $k$ and $j$
belonging to the same group and $2\overline{\varphi_{0}^{2}}$ otherwise) we
find that the parameter $r$ equals the length $l$ of the group. Therefore, we
can identify $\xi=rL_{0k}$ with the average domain size along the direction of
the cylinder axis, $L_{0k}$ is the distance between different sections of the
sample, see Fig.~\ref{Fig-1}. In Table~\ref{Table1} \begin{table}[h]
\caption{{}}%
\label{Table1}
\centering
\begin{tabular}
[c]{||c||c|c|c|c|c|c||}\hline\hline
Assembly ($^{\circ}$C) & \multicolumn{2}{c|}{A ($460$)} &
\multicolumn{2}{|c|}{B ($550$)} & \multicolumn{2}{|c||}{C ($640$)}\\\hline
Diameter (mm) & 6 & 8 & 6 & 8 & 6 & 8\\\hline\hline
$r^{\parallel}$ & 1.2 & 1.2 & 1.3 & 1.1 & 1.1 & 1.4\\\hline
$r^{\perp}$ & 1.6 & 1.4 & 1.4 & 1.3 & 1.3 & 1.5\\\hline\hline
\end{tabular}
\end{table}we present the results of calculations of the parameter $r$ for
specimens of parallel ($r^{\parallel}$) and perpendicular ($r^{\perp}$)
orientations averaged over all such samples.

Since the distance between different cross-sections of the specimen is
$L_{0k}=4.5$ mm, we get estimation $\xi\simeq6$ mm for the typical domain
size. As shown in Table~\ref{Table1}, specimens of perpendicular orientation
have higher $r$ than specimens of parallel orientation ($r^{\perp
}>r^{\parallel}$). The larger longitudinal size of domains with perpendicular
orientation is related to less restricted environment of such domains, that
weaker constraints the radiation-induced evolution of domain shape.

\begin{figure}
[tbh]
\begin{center}
\includegraphics[
height=1.9555in,
width=5.7655in
]%
{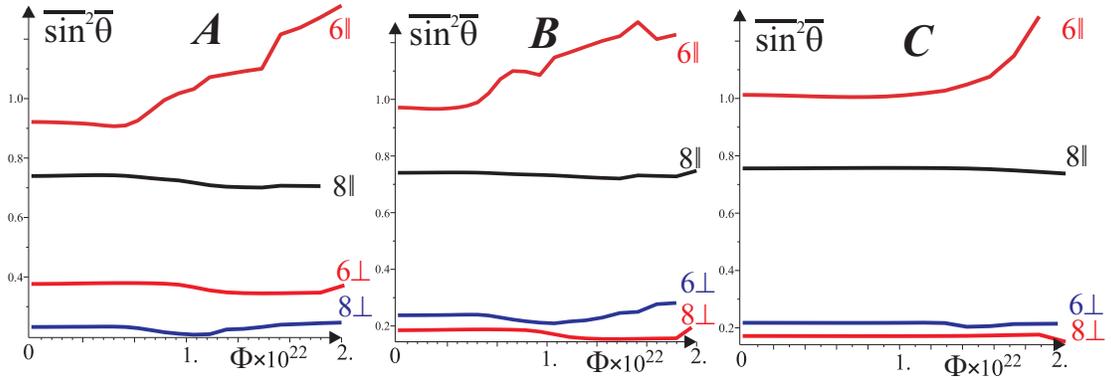}%
\caption{The dependence of $\overline{\sin^{2}\theta}$ on fluence $\Phi$ for
samples of assemblies $a=A,B,C$ with different average temperatures. Symbols
$d\omega$ correspond to samples of diameter $d=6$ and $8$ mm and orientation
$\omega$ ($\parallel$ and $\perp$).}%
\label{Angles}%
\end{center}
\end{figure}

The orientation of local axial axis with respect to the global axis of the
cylinder is described by the polar angle $\theta$ (see Fig. \ref{Fig-1} c).
Combining Eqs.~(\ref{eav}) and~(\ref{dll}) we find the average%
\begin{equation}
\overline{\sin^{2}\theta}=\frac{\bar{\varepsilon}}{\Delta d_{\max}/d-\Delta
L/L+\bar{\varepsilon}} \label{sin2}%
\end{equation}
Analyzing experimental data we show in Fig.~\ref{Angles} the dependence of
$\overline{\sin^{2}\theta}$ on fluence $\Phi$ for samples of different
diameters $d=6,8$ mm and orientations $\omega$ ($\parallel$ and $\perp$).%

One may ask whether the axial axis can be rotated as the result of radiation
induced shape-change of domains? Inspecting Fig.~\ref{Angles} we conclude that
this is not the case, and the orientation angle $\theta$ does not vary with
the fluence $\Phi$. The statistics of this angle only weakly depends on
temperature $T$ during irradiation, and thus is mainly determined by
fabrication conditions. An important special case are samples of diameter
$d=6$ mm with perpendicular orientation, that are strongly inhomogeneous and
have regions where radiation-induced shape-change is only weakly restricted by
the environment. Such heterogeneity of the sample is the reason of apparent
rise of $\overline{\sin^{2}\theta}$ above $1$ with the irradiation dose in
Fig.~\ref{Angles}. This effect is absent for samples of diameter $d=8$ mm,
supporting our previous estimation $\xi\simeq6$ mm for the domain size.

\begin{figure}
[tbh]
\begin{center}
\includegraphics[
height=2.6594in,
width=5.151in
]%
{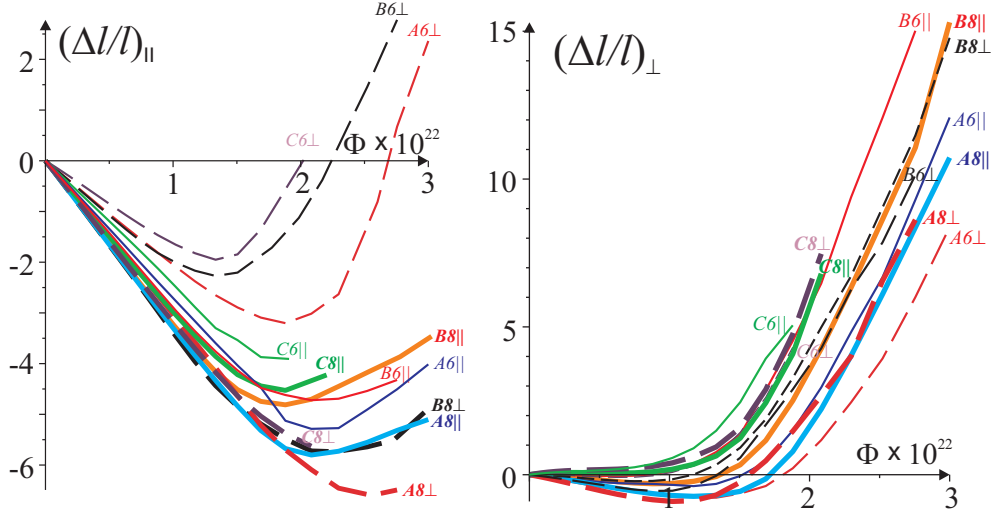}%
\caption{The dependence of principal strains $\left(  \Delta l/l\right)
_{\parallel}$ and $\left(  \Delta l/l\right)  _{\perp}$ (in \%) on neutron
fluence $\Phi$ for samples $(ad\omega)$ in the assembly $a=A,B,C$ with
diameter $d$ ($6$ mm - thin lines, $8$ mm - thick lines) and orientation
$\omega$ ($\parallel$ -- solid lines, $\perp$ -- dash lines). The same
notations as in Fig.~\ref{Length}.}%
\label{lpaer}%
\end{center}
\end{figure}

\emph{\textbf{d) Principal strains}}. The dependencies of principal strains
$\left(  \Delta l/l\right)  _{\parallel}$ and $\left(  \Delta l/l\right)
_{\perp}$ on neutron fluence $\Phi$ for all studied samples are collected in
Fig.~\ref{lpaer}.%
Below we use these data to discuss the dependence of these strains on shape
and size of samples, while the dose dependence will be considered later.

The dimension effect is the most pronounced for parallel principal strain
$\left(  \Delta l/l\right)  _{\parallel}$, see Fig.~\ref{lpaer} a). At small
fluence $\Phi$ the domain always shrinks in parallel direction. The amplitude
of the shrinking $\left(  \Delta l/l\right)  _{\parallel}$ is maximal for
specimens with perpendicular orientation, since domains of such samples are
less restricted to grow with increasing dose $\Phi$ with respect to specimens
of parallel orientation. This effect strongly depends on the sample size.
While the strain $\left(  \Delta l/l\right)  _{\parallel}$ dramatically
changes with orientation $\omega$ for samples of diameter $d=6$ mm, the value
$\left(  \Delta l/l\right)  _{\parallel}$ relatively weakly changes for
samples of diameter $d=8$ mm. This conclusion is in agreement with obtained
above estimation of the domain size $\xi\simeq6$ mm, and we expect that the
dimension effect is small only for samples of diameter $d>8$ mm.

The variation of $\left(  \Delta l/l\right)  _{\parallel}$ with the fluence
$\Phi$ is related to corresponding growth of microcracks which is hampered by
internal stresses in graphite. Such constrained effect of internal stresses is
minimal for specimens with perpendicular orientation. Similar behavior is
observed for perpendicular principal strain $\left(  \Delta l/l\right)
_{\perp}$, as well. At small $\Phi$ the value of $\left(  \Delta l/l\right)
_{\perp}$ weakly depends on fluence $\Phi$: The domain shrinks in
perpendicular direction at small temperature $T\lesssim600^{\circ}$ C and
extends at high temperatures $T>600^{\circ}$ C. The principal strain $\left(
\Delta l/l\right)  _{\perp}$ strongly varies with orientation $\omega$ only
for samples of diameter $d=6$ mm, and do not depend on orientation for samples
of diameter $d=8$ mm$.$ The amplitude of variation of $\left(  \Delta
l/l\right)  _{\perp}$ with neutron fluence $\Phi$ is maximal for the least
constrained in perpendicular principal direction specimens with parallel
orientation, and is minimal for specimens with perpendicular orientation.

The amplitude of anisotropy increases with the rise of the irradiation
temperature $T$. This effect is related to the development of an ensemble of
anisotropic microcracks and the rise of the microcrack volume with increasing
temperature $T$. The increase of the volume of microcracks with the rise of
the temperature is responsible for corresponding rise of the sample volume,
clearly observed in Fig.~\ref{Volume}. At high temperature ($T=650^{\circ}$ C)
the contribution of microcracks is so big, that at low doses it prevents the
shrinkage of the specimen length in parallel principal direction, $\left(
\Delta l/l\right)  _{\parallel}$, although there is always a region of initial
volume shrinking since at small $\Phi$ the domain shrinks in perpendicular
principal direction, $\left(  \Delta l/l\right)  _{\perp}$.

\begin{figure}
[tbh]
\begin{center}
\includegraphics[
height=2.4311in,
width=5.2638in
]%
{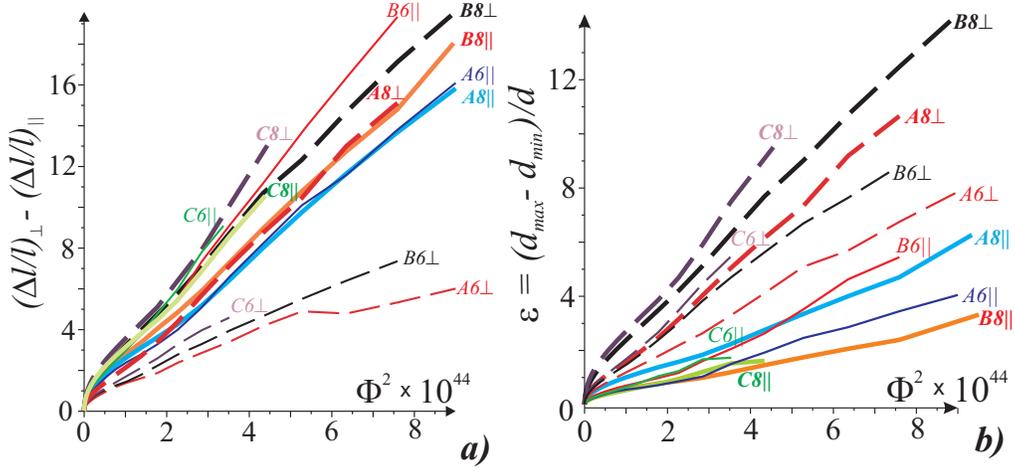}%
\caption{The dependence (in \%) of the principal strain difference $\left(
\Delta l/l\right)  _{\perp}-\left(  \Delta l/l\right)  _{\parallel}$ a) and of
the flattening factor $\bar{\varepsilon}$ b) on squared fluence $\Phi$ for all
samples. The same notations as in Fig.~\ref{Length}.}%
\label{Delta}%
\end{center}
\end{figure}

Principal strains $\left(  \Delta l/l\right)  _{\parallel}$ and $\left(
\Delta l/l\right)  _{\perp}$ demonstrate essentially non-monotonic dependence
on fluence $\Phi$, see Fig. \ref{lpaer}. In order to understand the reason of
such behavior we can present these strains in the form%
\begin{align}
\left(  \frac{\Delta l}{l}\right)  _{\parallel}  &  =\frac{1}{3}\frac{\Delta
V}{V}-\frac{2}{3}\left[  \left(  \frac{\Delta l}{l}\right)  _{\perp}-\left(
\frac{\Delta l}{l}\right)  _{\parallel}\right]  ,\label{dl1}\\
\left(  \frac{\Delta l}{l}\right)  _{\perp}  &  =\frac{1}{3}\frac{\Delta V}%
{V}+\frac{1}{3}\left[  \left(  \frac{\Delta l}{l}\right)  _{\perp}-\left(
\frac{\Delta l}{l}\right)  _{\parallel}\right]  \label{dl2}%
\end{align}
The non-monotonic dependence of these strains on $\Phi$ is related to
corresponding non-monotonic dependence of the volume, see Fig.~\ref{Volume}.
The dependence of the difference of principal strains, $\left(  \Delta
l/l\right)  _{\perp}-\left(  \Delta l/l\right)  _{\parallel}$, on fluence
$\Phi$ is monotonic and can be extrapolated by quadratic function, see
Fig.~\ref{Delta} a:%
\begin{equation}
\left(  \Delta l/l\right)  _{\perp}-\left(  \Delta l/l\right)  _{\parallel
}\simeq\beta_{l}\Phi^{2} \label{DeltaF}%
\end{equation}%
The shape of samples becomes more and more crankshaft-like with the rise of
the neutron fluence $\Phi$. The amplitude of this effect is characterized by
the average flattening factor $\bar{\varepsilon}$ of the elliptic
cross-sections of the specimen, Eq.~(\ref{eav}). Inspecting experimental data
we find that $\bar{\varepsilon}$ grows quadratically with the fluence $\Phi$:%
\begin{equation}
\bar{\varepsilon}\left(  \Phi\right)  \simeq\beta_{\varepsilon}\Phi^{2}
\label{epsf}%
\end{equation}
The dependence~(\ref{epsf}) is illustrated in Fig.~\ref{Delta} b), where we
plot $\bar{\varepsilon}$ for different assembles, diameters and orientations
of specimens as function of $\Phi^{2}$.

The coefficients $\beta_{l}$ and $\beta_{\varepsilon}$ for different
orientations $\omega$, diameters $d$ and assembles $a$ are shown in
Table~\ref{Table2}. \begin{table}[h]
\label{Table2}\centering
\begin{tabular}
[c]{||c||c|c|c|c|c|c||}\hline\hline
Diameter $d$ & \multicolumn{3}{c|}{6 mm} & \multicolumn{3}{c||}{8 mm}\\\hline
Assembly $a$ & A & B & C & A & B & C\\\hline\hline
$\left(  \beta_{l}\right)  ^{\parallel}\times10^{-46}$ & 1.8 & 2.5 & 2.5 &
1.7 & 2.0 & 2.3\\\hline
$\left(  \beta_{l}\right)  ^{\perp}\times10^{-46}$ & 0.7 & 1.0 & 1.2 & 2.0 &
2.2 & 2.8\\\hline\hline
$\left(  \beta_{\varepsilon}\right)  ^{\parallel}\times10^{-46}$ & 0.4 & 0.7 &
0.5 & 0.7 & 0.4 & 0.4\\\hline
$\left(  \beta_{\varepsilon}\right)  ^{\perp}\times10^{-46}$ & 0.9 & 1.2 &
1.5 & 1.4 & 1.6 & 2.0\\\hline\hline
\end{tabular}
\end{table}The factor $\beta_{l}$ monotonically grows with the temperature $T$
during irradiation (from assembly $A$ to $C$). The difference of $\beta_{l}$
for specimens of parallel, $\left(  \beta_{l}\right)  ^{\parallel}$, and
perpendicular, $\left(  \beta_{l}\right)  ^{\perp}$, orientations decreases
with the rise of the sample diameter, but the gap between these values exists
even for maximal diameter $d=8$ mm. Therefore, we conclude that in our
experiment even the most thick samples have elastic properties different from
properties of a bulk graphite.

The coefficients $\beta_{\varepsilon}$ increase with the rise of the
temperature $T$. According to Eq.~(\ref{eav}) the two coefficients $\beta_{l}$
and $\beta_{\varepsilon}$ are proportional to each other:
\begin{equation}
\beta_{\varepsilon}=\overline{\sin^{2}\theta}\beta_{l}%
\end{equation}
Higher values of $\beta_{\varepsilon}$ for specimens of perpendicular
orientation with $\omega=\pi/2$ are related to larger values $\overline
{\sin^{2}\theta}$ with respect to the case of parallel orientation and
orientation angle $\omega=0$. We expect that for bulk graphite both
coefficients vanish, $\beta_{l}=\beta_{\varepsilon}=0$, and there should not
be any difference between the two ways of measurement of the volume change,
$\Delta V^{C}=\Delta V^{F}$, Eqs.~(\ref{Vt}) and~(\ref{Vprime}).

At first sight simple quadratic dependence~(\ref{DeltaF}) looks very
surprising, taking into account complex non-monotonic dependence of volume
changes on the fluence $\Phi$, see Fig.~\ref{Volume}: at low dose $\Phi
<\Phi_{\min}$ the sample is shrinking, while at high doses it dilates. It is
usually assumed that appearance at large fluences $\Phi>\Phi_{\min}$ of
growing branches in volume and elongation dependences is related to formation
of new set of microcracks in graphite at moderate fluences $\Phi>\Phi_{\min}%
$.\cite{1} Simple dependence~(\ref{DeltaF}) demonstrates that really nothing
special is happened with graphite above the cross-over $\Phi_{\min}$. In Ref.
\cite{1} we proposed alternative explanation of such non-monotonic dependences
as the result of the interplay of radiation induced changes of the shape and
variations of elastic moduli of crystallites in polycrystal graphite. At low
doses, $\Phi<\Phi_{\min}$, the change of the shape of domains wins, leading to
initial shrinking of graphite samples. At high doses, $\Phi>\Phi_{\min}$, the
main contribution to graphite deformation comes from the variation of elastic
moduli, leading to dilation of graphite.

\begin{figure}
[tbh]
\begin{center}
\includegraphics[
height=5.2485in,
width=5.0662in
]%
{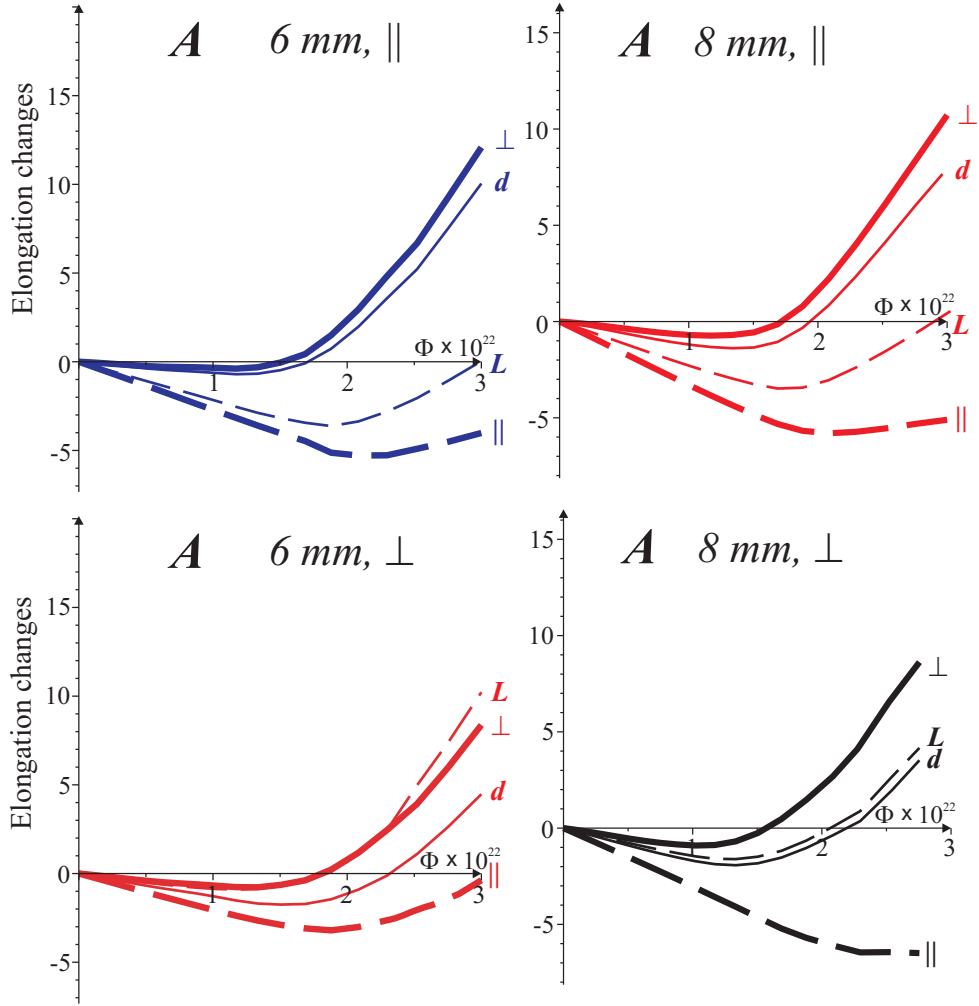}%
\caption{The dependence of relative elongations (in \%) of samples in reactor
BOR-60: $\Delta L/L$ (dashed thin lines, symbol $L$), $\Delta d/d$ (solid thin
lines, symbol $d$), $\left(  \Delta l/l\right)  _{\perp}$ (solid thick lines,
$\perp$) and $\left(  \Delta l/l\right)  _{\parallel}$ (dashed thick lines,
$\parallel$) on fluence $\Phi$ for samples of the assembly $A$ with average
irradiation temperature $T=450^{\circ}$C$.$ The same notations as in
Fig.~\ref{Length}.}%
\label{A-1}%
\end{center}
\end{figure}

\emph{\textbf{e) Relative contributions}}. In Fig.~\ref{A-1} we show the
relative contribution of four important geometrical characteristics of samples
of assembly $A$: the change of length $L$, the change of diameter $d$ and two
principal strains $\left(  \Delta l/l\right)  _{\parallel}$ and $\left(
\Delta l/l\right)  _{\perp}$ as functions of fluence $\Phi$.%
Due to Eq.~(\ref{Vt}) there is linear dependence between these variables. We
observe that measured elongations $L$ and $d$ lie in between principal
elongations $\left(  \Delta l/l\right)  _{\parallel}\ $and $\left(  \Delta
l/l\right)  _{\perp}$ because of constraints imposed on the domain shape in
the volume of graphite.

\section{Theory: Internal stresses in graphite}

\begin{figure}
[tbh]
\begin{center}
\includegraphics[
height=5.3116in,
width=3.5014in
]%
{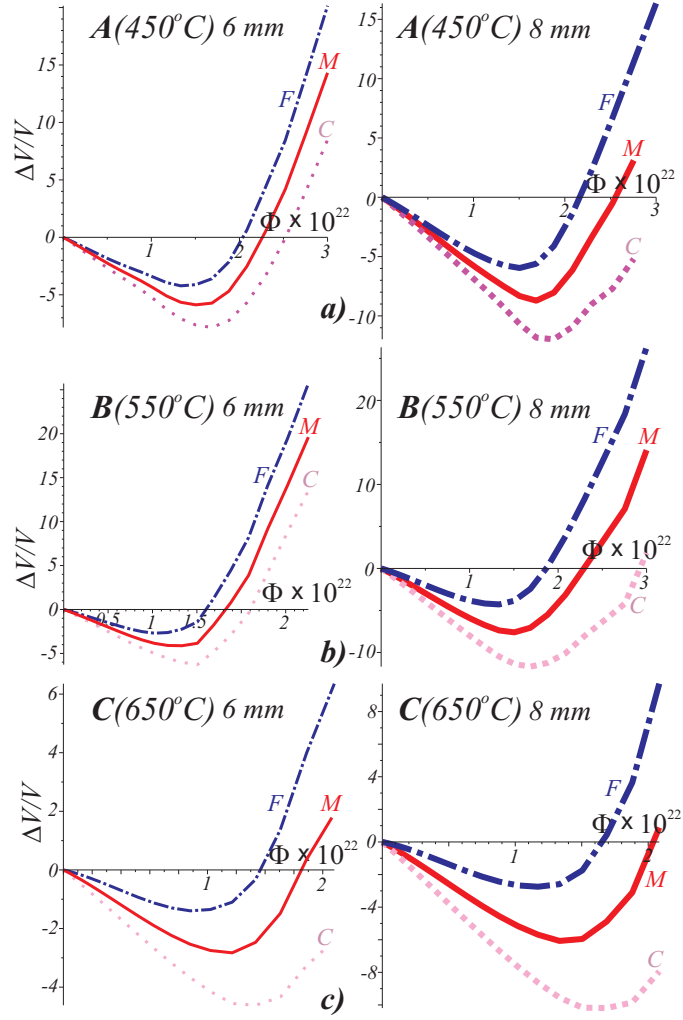}%
\caption{The dependence of relative volume changes (in \%) for samples of
diameters $d=6$ and $8$ mm on fluence $\Phi$ for assemblies $a=A,B,C$,
corresponding to irradiation temperatures $T=460,550$ and $640^{\circ}$C:
upper dash-dotted lines $F$ give actual values, $\left(  \Delta V/V\right)
^{F}$, Eq.~(\ref{dVF}), lower dotted lines $C$ correspond to traditional
expressions, $\left(  \Delta V/V\right)  ^{C}$, Eq.~(\ref{dV/V}), and middle
solid lines $M$ are our predictions for bulk graphite, $\left(  \Delta
V/V\right)  ^{M}$, Eq.~(\ref{dVMS}).}%
\label{FCM}%
\end{center}
\end{figure}

Intuitively, when all domains are strictly oriented, the radiation-induced
change of their shape does not cause any internal stresses. According to the
terminology of Ref.\cite{n+8} such domains are in the state of a free
dilatation. In this state the relative volume change of a sample of arbitrary
shape and of the macro-graphite are the same. This idealized case corresponds
to expression (\ref{dV/V}) and the upper curves in Figs. \ref{FCM}, in the
sense that they describe the change of domain shape in almost no stress,
despite of apparent disorientation of domains. In this case internal stresses
are absent because the size of domains is comparable with the minimum size
(the diameter) of the samples. In the real case of macro-graphite with all
dimensions much larger than the domain size, the situation is fundamentally
different. Below, we consider this case and describe the emerging internal
stresses because of domain disorientation (see Fig.~\ref{Domains} b) following
the method developed in Ref.\cite{n+8}.%

Let us analyze the meaning of differences in the volume change $(\Delta V/V)$
obtained in Eqs.~(\ref{dVV}) and~(\ref{dV/V}). Expression~(\ref{dVV})
corresponds, as already mentioned, to the radiation-induced \textquotedblleft
free dilatation\textquotedblright. Now turn to Eq.~(\ref{dV/V}). Longitudinal
elongation in this expression (for example, $\left(  \Delta L/L\right)
_{\parallel}$ for parallel orientation) describes the real strain of the
sample, taking into account the disorientation of its domains. Transversal
elongation ($\left(  \Delta L/L\right)  _{\perp}$ for parallel orientation) in
expression~(\ref{dV/V}) corresponds to totally suppressed deformations of
adjacent domains of distinct orientations. Eq.~(\ref{dV/V}) gives the relative
volume change of a simply connected macro-graphite assembled from the samples
of say, parallel orientation, when all discrepancies between adjacent
shape-changing domains are straighten by stresses at domain boundaries. The
difference in values~(\ref{dVV}) and~(\ref{dV/V}) determines the excess volume
fraction of such overlapping radiation-induced regions:%
\begin{equation}
\delta_{0}\left(  \frac{\Delta V}{V}\right)  \equiv\left(  \frac{\Delta V}%
{V}\right)  ^{F}-\left(  \frac{\Delta V}{V}\right)  ^{C}=2\bar{\varepsilon
}^{\perp}-\bar{\varepsilon}^{\parallel} \label{dVFS}%
\end{equation}
After reaching the equilibrium at internal boundaries between the domains the
excess volume fraction $\delta_{0}(\Delta V/V)$ partially relaxes. Considering
a macro-graphite as a body, containing an ensemble of equilibrium dilatating
objects whose dimensions are much smaller than the size of the body with free
boundaries, one can find the dilatation component of the total strain.
Following Ref.\cite{n+8} we get in the leading order in $(\Delta V/V)$:%
\begin{equation}
Tr\left(  \hat{u}\right)  \ \mathbf{\simeq}\frac{2}{3}\frac{1-2\nu}{1-\nu
}\left(  2\bar{\varepsilon}^{\perp}-\bar{\varepsilon}^{\parallel}\right)
\label{Tru}%
\end{equation}
Since in the reference state the relative volume change is $(\Delta V/V)^{C}$,
we find expression for the relative change of the volume of macro-graphite:%
\begin{align}
\left(  \Delta V/V\right)  ^{M}  &  =\left(  \Delta V/V\right)  ^{C}+Tr\left(
\hat{u}\right) \nonumber\\
&  \mathbf{\simeq}\left(  \frac{\Delta L}{L}\right)  ^{\parallel}+2\left(
\frac{\Delta L}{L}\right)  ^{\perp}+\frac{2}{3}\frac{1-2\nu}{1-\nu}\left(
2\bar{\varepsilon}^{\perp}-\bar{\varepsilon}^{\parallel}\right)  \label{dVMS}%
\end{align}
The relation between different definitions of relative volume change, $(\Delta
V/V)^{F}$, $(\Delta V/V)^{C}$ and $(\Delta V/V)^{M}$, is shown in
Figs.~\ref{FCM} taking into account the value of Poisson's ratio $\nu=0.2$ for
graphite GR-280.

The obtained result is in agreement with frequently noted in the literature
experimental fact -- that large samples of extruded graphite change their
shape stronger than the thinner samples. For example, the paper \cite{n+11}
presents the results of measurements of irradiated samples, $\sim10$ cm
$\times$ $10$ cm $\times$ $61$ cm, and also of cylindric specimens with
diameter $1.09$ cm and length $10$ cm, in the case of parallel and
perpendicular orientations with respect to the extrusion axis. The observed
values of the volume change (the shrinkage) for larger samples are
approximately twice as large. Using expressions~(\ref{dVFS}) and~(\ref{dVMS})
the relation between macro-graphite and thin samples can be written as%
\begin{equation}
\left(  \frac{\Delta V}{V}\right)  ^{M}=\left(  \frac{\Delta V}{V}\right)
^{F}-\frac{1}{3}\frac{1+\nu}{1-\nu}\left(  2\bar{\varepsilon}^{\perp}%
-\bar{\varepsilon}^{\parallel}\right)  \label{dVMF}%
\end{equation}

\section{Conclusions}

In this work we present data on both longitudinal and transverse measurements
of radiation-induced shape-change for graphite specimens of cylindrical shape.
These data show that:

\begin{itemize}
\item The cross-sections of the specimens change their shape under irradiation
from initial circular to elliptical one, and orientations of main ellipse axes
randomly vary along the sample.

\item The relative volume changes, obtained by direct measurement of $\Delta
V/V^{F}$ and using the traditional expression $\left(  \Delta V/V\right)
^{C}$ in terms of relative length variations of samples with parallel and
perpendicular orientations, are essentially different (see
expression~(\ref{Vprime}) and Fig.~\ref{FCM}).
\end{itemize}

To explain these facts, authors proposed a new model of the morphology of
extruded graphite, considering it as an ensemble of dilated domains with
orientations disoriented with respect to the global axis of symmetry
(direction of extrusion) of macro-graphite. In this model the symmetry
properties of both macro-graphite and domains are the same -- the transversal isotropy.

For each cross-section of cylindrical specimens of diameters 6 and 8 mm with
parallel and perpendicular orientations we present results of measured values
$\left(  \Delta L/L\right)  ^{\parallel},\left(  \Delta L/L\right)  ^{\perp}$,
$d_{\max}$ and $d_{\min}$. We derive closed system of equations for parameters
$\left(  \Delta l/l\right)  _{\parallel},\left(  \Delta l/l\right)  _{\perp}$
and $\overline{\sin^{2}\theta}$ characterizing deformation of domains and
their orientation and calculate these values.

Our model predicts that domains of samples with transverse dimensions on the
order of domain size should experience free (unconstrained) radiation-induced
dilation. In this case, the values $\left(  \Delta l/l\right)  _{\parallel
},\left(  \Delta l/l\right)  _{\perp}$ and $\overline{\sin^{2}\theta}$ must
match for samples with parallel and perpendicular orientations. This condition
is satisfied for specimens of diameter 6 mm and 8 mm (see Fig.~\ref{A-1}). The
data for such free-dilatating domains were used in the subsequent analysis and evaluations.

The main results can be summarized as:

\begin{enumerate}
\item[$\alpha.$] In macro-graphite radiation-induced dilation of disoriented
($\overline{\sin^{2}\theta}>0$, see Fig.~\ref{Domains} b) domains leads to the
development of cross-domain stresses. Of particular interest is the dilatation
component of the stress leading to the difference between real volume change
$\left(  \Delta V/V\right)  ^{F}$ and commonly used expression $\left(  \Delta
V/V\right)  ^{C}$ for this change. Note that in the case of completely ordered
domains (see Fig.~\ref{Domains} a) internal stresses disappear and all
expressions for the volume change are the same,%
\begin{equation}
\left(  \frac{\Delta V}{V}\right)  ^{F}=\left(  \frac{\Delta V}{V}\right)
^{C}=\left(  \frac{\Delta V}{V}\right)  ^{M}%
\end{equation}

\item[$\beta.$] It is shown that the shrinkage curves obtained both by direct
measurements and by calculation of relative size changes do not describe the
real shrinkage curve of macro-graphite. We derived the relation%
\begin{equation}
\left(  \frac{\Delta V}{V}\right)  ^{M}=\left(  \frac{\Delta L}{L}\right)
^{\parallel}+2\left(  \frac{\Delta L}{L}\right)  ^{\perp}+\frac{2}{3}%
\frac{1-2\nu}{1-\nu}\left(  2\bar{\varepsilon}^{\perp}-\bar{\varepsilon
}^{\parallel}\right)  \label{VM}%
\end{equation}
that allows to obtain the shrinkage curve of macro-graphite by direct
measurements of strains in samples of small sizes.

\item[$\gamma.$] Our analysis leads to important conclusion about the presence
of size restrictions on experimental specimens. In the case when minimum
sample size $L_{\min}$ is much larger than the domain size $\xi$, $L_{\min}%
\gg\xi$, the elongation measurements directly match the behavior of
macroscopic graphite. However, such a condition is very hard to implement in
reactor experiments.
\end{enumerate}

The simplest realized condition is to satisfy the ratio $L_{\min}\simeq\xi$,
with subsequent conversion of obtained data using the relation~(\ref{VM}). In
our experiments, as seen from Fig.~\ref{A-1}, the condition $L_{\min}\simeq
\xi$ is realized for the set of specimens with diameters of $6$ mm and $8$ mm.
This criterion is in a satisfactory agreement between principal elongations
$\left(  \Delta l/l\right)  _{\parallel}$ and $\left(  \Delta l/l\right)
_{\perp}$ of dilatating domains of different orientations.

\begin{acknowledgments}
The authors consider as their pleasant duty to thank B.N. Bennesch, A.N.
Maltsev and A.V. Nechitailo who made a significant contribution to the experiments.
\end{acknowledgments}

\appendix{\Large \textbf{Appendix}}%

\def\theequation{A\arabic{equation}}
\setcounter{equation}{0}%

\section*{Solution of minimum conditions}

The angular dependence of the diameter~(\ref{dphi}) of deformed cylinder with
initial circular cross-section can be rewritten in the form%
\begin{equation}
d^{2}\left(  \varphi\right)  =a_{k}+b_{k}\cos\left(  2\varphi\right)
+c_{k}\sin\left(  2\varphi\right)  \label{d2f}%
\end{equation}
where%
\begin{align}
a_{k}  &  =\frac{1}{2}\left[  d_{\max}^{2}+\left(  d_{\min}^{2}\right)
_{k}\right]  ,\nonumber\\
b_{k}  &  =\frac{1}{2}\left[  d_{\max}^{2}-\left(  d_{\min}^{2}\right)
_{k}\right]  \cos\left[  2\left(  \varphi_{0}\right)  _{k}\right]
,\label{abc}\\
c_{k}  &  =\frac{1}{2}\left[  d_{\max}^{2}-\left(  d_{\min}^{2}\right)
_{k}\right]  \sin\left[  2\left(  \varphi_{0}\right)  _{k}\right] \nonumber
\end{align}
Knowing coefficients $a_{k},b_{k}$ and $c_{k}$ we can calculate from these
equations the two principal diameters $d_{\max}$ and $\left(  d_{\min}\right)
_{k}$:%
\begin{align}
d_{\max}^{2}  &  =\frac{1}{N}\sum_{k=1}^{N}\left\{  a_{k}+b_{k}\cos\left[
2\left(  \varphi_{0}\right)  _{k}\right]  +c_{k}\sin\left[  2\left(
\varphi_{0}\right)  _{k}\right]  \right\}  ,\label{dd}\\
\left(  d_{\min}^{2}\right)  _{k}  &  =2a_{k}-d_{\max}^{2}\nonumber
\end{align}

In order to find coefficients $a_{k},b_{k}$ and $c_{k}$ from experimental data
we substitute Eq.~(\ref{d2f}) into Eq.~(\ref{dphi}). The result of this
substitution can be simplified using normalization conditions%
\begin{equation}
\sum_{i=1}^{M}\cos^{2}\left(  2\varphi_{i}\right)  =\sum_{i=1}^{M}\sin
^{2}\left(  2\varphi_{i}\right)  =\frac{\allowbreak M}{2}%
\end{equation}
and orthogonality relations%
\begin{equation}
\sum_{i=1}^{M}\cos\left(  2\varphi_{i}\right)  \sin\left(  2\varphi
_{i}\right)  =\sum_{i=1}^{M}\cos\left(  2\varphi_{i}\right)  =\sum_{i=1}%
^{M}\sin\left(  2\varphi_{i}\right)  =0
\end{equation}
After some algebra we get%
\begin{equation}
\sigma^{2}=M\sum_{k=1}^{N}\left[  2a_{k}^{2}+b_{k}^{2}+c_{k}^{2}-4a_{k}%
A_{k}-2b_{k}B_{k}-2c_{k}C_{k}\right]  , \label{sigma}%
\end{equation}
where $A_{k},B_{k}$ and $C_{k}$ are defined in Eq.~(\ref{Dmax}) --~(\ref{phik}%
). Minimizing expression~(\ref{sigma}) with respect to $\left(  d_{\min
}\right)  _{k}$, $d_{\max}$ and $\left(  \varphi_{0}\right)  _{k}$ we find
equations for these variables%
\begin{align}
2\left(  a_{k}-A_{k}\right)  -\left(  b_{k}-B_{k}\right)  \cos\left[  2\left(
\varphi_{0}\right)  _{k}\right]  -\left(  c_{k}-C_{k}\right)  \sin\left[
2\left(  \varphi_{0}\right)  _{k}\right]   &  =0,\nonumber\\
2\sum\nolimits_{k=1}^{N}\left\{  \left(  a_{k}-A_{k}\right)  +\left(
b_{k}-B_{k}\right)  \cos\left[  2\left(  \varphi_{0}\right)  _{k}\right]
+\left(  c_{k}-C_{k}\right)  \sin\left[  2\left(  \varphi_{0}\right)
_{k}\right]  \right\}   &  =0\label{aA}\\
-\left(  b_{k}-B_{k}\right)  \sin\left[  2\left(  \varphi_{0}\right)
_{k}\right]  +\left(  c_{k}-C_{k}\right)  \cos\left[  2\left(  \varphi
_{0}\right)  _{k}\right]   &  =0\nonumber
\end{align}
Inspection of the above equations shows that $b_{k}$ and $c_{k}$ can be
written in the form%
\begin{equation}
b_{k}=\sqrt{b_{k}^{2}+c_{k}^{2}}\cos\left[  2\left(  \varphi_{0}\right)
_{k}\right]  ,\qquad c_{k}=\sqrt{b_{k}^{2}+c_{k}^{2}}\sin\left[  2\left(
\varphi_{0}\right)  _{k}\right]  \label{bc}%
\end{equation}
where $\left(  \varphi_{0}\right)  _{k}$ is the solution of equation%
\begin{equation}
B_{k}\sin\left[  2\left(  \varphi_{0}\right)  _{k}\right]  =C_{k}\cos\left[
2\left(  \varphi_{0}\right)  _{k}\right]  \label{phiAB}%
\end{equation}

Substituting Eqs.~(\ref{bc}) and~(\ref{phiAB}) into Eq.~(\ref{aA}) we find%
\begin{align}
2\left(  a_{k}-A_{k}\right)  -\sqrt{b_{k}^{2}+c_{k}^{2}}+\sqrt{B_{k}^{2}%
+C_{k}^{2}}  &  =0,\\
2\sum\nolimits_{k=1}^{N}\left\{  \left(  a_{k}-A_{k}\right)  +\sqrt{b_{k}%
^{2}+c_{k}^{2}}-\sqrt{B_{k}^{2}+C_{k}^{2}}\right\}   &  =0
\end{align}
Combining these equations we get%
\begin{equation}
b_{k}^{2}+c_{k}^{2}=B_{k}^{2}+C_{k}^{2} \label{bB}%
\end{equation}
and also%
\begin{equation}
a_{k}=A_{k}+\sqrt{B_{k}^{2}+C_{k}^{2}} \label{ak}%
\end{equation}
Substituting Eq.~(\ref{bB}) into Eq.~(\ref{bc}) we find coefficients $b_{k}$
and $c_{k}$:%
\begin{equation}
b_{k}=\sqrt{B_{k}^{2}+C_{k}^{2}}\cos\left[  2\left(  \varphi_{0}\right)
_{k}\right]  ,\qquad c_{k}=\sqrt{B_{k}^{2}+C_{k}^{2}}\sin\left[  2\left(
\varphi_{0}\right)  _{k}\right]  \label{bkck}%
\end{equation}

As the last step of our derivation we substitute Eqs.~(\ref{ak})
and~(\ref{bkck}) into Eqs.~(\ref{dd}) and find diameters $d_{\max}$ and
$\left(  d_{\min}\right)  _{k}$, Eqs.~(\ref{Dmax}) and~(\ref{Dmin}).

\end{document}